\documentclass[a4,10pt,onecolumn]{article}
\usepackage{graphicx} 
\usepackage[margin=0.8in]{geometry} 
\usepackage{amsmath,amsthm,amssymb}
\usepackage{float}
\usepackage{xcolor}
\usepackage[utf8]{inputenc}
\usepackage{graphicx}
\usepackage[backend=bibtex,style=nature]{biblatex}
\usepackage{lineno}
\usepackage{authblk}
\usepackage{subcaption}
\usepackage{tabularx}
\usepackage{booktabs}
\usepackage{makecell}
\usepackage{colortbl}
\usepackage{multirow}
\graphicspath{{./Figures/}} 
\definecolor{rowblue}{RGB}{200,215,223}
\definecolor{sideblue}{RGB}{18,96,132}

\def\_#1{{\bf #1\mit}}

\usepackage[font=footnotesize]{caption}

\title{Boosting Sensing Performance through Near-Field Engineering in Low-Q Metasurfaces}

\date{}
\author[1]{J.A. Álvarez-Sanchis}
\author[1]{L.M. Máñez-Espina}
\author[1]{T. Mengual Chuliá}
\author[1]{A. Griol}
\author[1]{A. Díaz-Rubio$^*$}
\affil[1]{Nanophotonics Technology Center, Universitat Politècnica de València, València 46022,  Spain.}
\affil[*]{andiaru@upv.es}

\begin{document}

\maketitle

\begin{abstract} 
Dielectric metasurfaces have introduced a new paradigm for substance detection by exploiting their resonant properties to enhance light–matter interaction. This enhancement can be used for sensing either through refractive index changes or through absorption-based mechanisms. Most works focus on high–quality factor resonators, aiming to increase field confinement in the vicinity of the resonant structure to improve sensitivity. In this work, we explore an alternative approach based on low–quality factor, fully dielectric metasurfaces, with engineered modes to enhance near-field concentration. We investigate different topologies that, despite their low quality factors, achieve sensitivity and detection performance beyond what is typically reported for low–Q structures in the literature. This improvement is enabled by near-field engineering of the evanescent modes, allowing us to control the spatial distribution of the electromagnetic field and maximize its overlap with the analyte.
Our results show that careful mode engineering provides a powerful strategy to boost sensing performance without relying on ultra-high-Q resonances.

\end{abstract}

\section{Introduction}

Enhancing light--matter interaction at the nanoscale is a central objective in modern photonics, particularly for chemical and biological sensing applications where weak perturbations of the surrounding environment must be detected with high precision. Metasurfaces have emerged as a powerful platform to address this challenge due to their ability to manipulate electromagnetic waves using subwavelength resonant elements. The optical response of these planar nanostructures is governed by the geometry of their meta-atoms, enabling precise control over the amplitude, phase, and polarization of scattered fields. When illuminated, metasurfaces support resonant modes that strongly confine electromagnetic energy in the near field, thereby enhancing the interaction with surrounding analytes. Owing to these properties, metasurfaces have been extensively explored for refractive-index sensing, chemical detection, and biodetection, where sensing performance relies on efficient coupling between optical fields and the target medium~\cite{beruete2020terahertz,john2023metasurface}.

In resonant nanophotonic sensing platforms, light--matter interaction is mediated by resonant modes supported by the metasurface. These resonances can be classified according to their quality factor $Q$ into low ($10$--$100$), high ($10^2$--$10^3$), and ultra-high ($10^3$--$10^6$ or higher) regimes. High- and ultra-high-$Q$ resonances exhibit narrow spectral linewidths and long photon lifetimes, leading to enhanced spectral resolution and large figures of merit defined as $\mathrm{FOM}=S/\Delta\lambda$, where $S$ is the refractive-index sensitivity and $\Delta\lambda$ is the resonance linewidth. These properties make them attractive for detecting small refractive-index variations~\cite{yu2025free}. Consequently, considerable effort has been devoted to engineering metasurfaces that support high-$Q$ responses, including guided-mode resonances, surface lattice resonances, and quasi-bound states in the continuum, where destructive interference or symmetry protection suppresses radiative losses.
Despite these advantages, maintaining high-$Q$ resonances in practical sensing environments remains challenging. Their performance is highly sensitive to fabrication imperfections, material absorption~\cite{alvarez2023loss}, and finite-size effects, while their weak radiative coupling complicates experimental excitation. In sensing applications, additional absorption introduced by the analyte can broaden the resonance and reduce the effective quality factor. 

Resonances with moderate or low $Q$ factors offer several practical advantages in this context. In particular, they typically exhibit stronger coupling to free space, larger resonance amplitudes, and increased robustness against fabrication imperfections and environmental losses. Their shorter photon lifetimes also enable faster temporal responses, which can be beneficial in dynamic sensing environments such as biological media. However, the broader linewidth associated with low-$Q$ resonances generally reduces the achievable figure of merit, meaning that alternative design strategies are required to achieve high sensing performance.
Plasmonic metasurfaces supporting localized surface plasmon resonances (LSPRs) constitute a well-known example of sensing platforms operating in the low-$Q$ regime~\cite{mayer2011localized,nieves2023development}. These structures provide strong electromagnetic field confinement near the metal surface, resulting in high sensitivity to refractive-index variations. However, their performance is fundamentally limited by intrinsic Ohmic losses in metals, which broaden the resonance and reduce the spectral contrast~\cite{conteduca2022beyond}. Since the limit of detection can be approximated as $LOD \propto \lambda_0/(S\cdot Q_{\rm R} \cdot A)$, where $A$ represents the resonance amplitude, and $Q_{\rm R}$ is the radiative quality factor. Therefore, a reduction in resonance amplitude directly reduces the detectability capacity.

An attractive alternative consists of employing all-dielectric metasurfaces that avoid metallic absorption while preserving strong light--matter interaction. In such systems, moderate-$Q$ resonances can maintain large spectral amplitudes due to the low intrinsic losses of dielectric materials. However, dielectric resonances often confine a significant fraction of the electromagnetic energy inside the nanostructures~\cite{yesilkoy2019ultrasensitive, ren2024overcoming}, which reduces the overlap between the resonant field and the surrounding analyte and therefore limits the achievable sensitivity. 
From a theoretical perspective, the sensitivity of a resonant nanophotonic sensor is governed by the overlap between the electromagnetic field and the analyte region. It is found that the sensitivity of a resonant structure is directly proportional to the fraction of the field in the analyte region compared to the dielectric region of interest ~\cite{zalyubovskiy2012theoretical}:
\begin{equation}\label{eq:1}
f=
\frac{\displaystyle \int_{V_{\rm a}} \varepsilon_{\rm r}(\mathbf{r}) |E(\mathbf{r})|^2 \, dV}
{ \displaystyle\int_{V_{\rm d}} \varepsilon_{\rm r}(\mathbf{r}) |E(\mathbf{r})|^2 \, dV},
\end{equation}
where $V_\mathrm{a}$ denotes the sensing volume occupied by the analyte, $V_{\rm d}$ is the near-field volume,  and $\varepsilon_{\rm r}(\mathbf{r})$ represents the spatial relative permittivity distribution. In Equation (\ref{eq:1}), $|E(\mathbf{r})|^2=E(\mathbf{r})\cdot E^{*}(\mathbf{r})$, where the symbol $^*$ denotes complex conjugation. Indeed, $S \propto f\cdot Q$. This expression relates the achievable sensitivity to the fraction of electromagnetic energy located within the sensing region rather than on the quality factor alone. Consequently, maximizing the near-field overlap with the analyte becomes a key design strategy when working with moderate or low-$Q$ resonances.

Several studies have explored near-field engineering to improve the sensing performance of nanophotonic resonators. In high-$Q$ metasurfaces, such as those based on quasi-bound states in the continuum or guided-mode resonances, different approaches have been proposed to redistribute the modal field toward the surrounding medium and enhance the interaction with the analyte~\cite{li2023balancing,li2024homogeneous}. Similarly, in plasmonic platforms, near-field engineering has been widely used to increase field confinement in the sensing region despite the relatively low quality factors associated with Ohmic losses~\cite{chae2016engineering, ren2024overcoming}. However, analogous strategies remain largely unexplored in low-loss dielectric metasurfaces operating in the moderate- or low-$Q$ regime, particularly in systems based on localized Mie resonances, where the electromagnetic field is predominantly confined inside the dielectric nanostructure. 
Recent studies have shown that dielectric resonators can be engineered to confine the electromagnetic field in low-index regions such as air rather than inside the high-index dielectric material~\cite{hentschel2023dielectric,ludescher2025optical}. While this approach increases the accessibility of the modal field for sensing, the resulting spectral response is often weak, exhibiting limited variations in reflection or scattering intensity, which reduces the achievable spectral contrast and limits their effectiveness for refractive-index sensing.

In this work, we follow this approach and perform modal engineering of low-$Q$ dipolar resonances in dielectric metasurfaces and mode hybridization to enhance field concentration in the analyte volume. In Section 2, we show how to redistribute the electromagnetic near field toward the external sensing region by tailoring the geometry of the meta-atoms, maximizing the interaction between the resonant mode and the surrounding analyte. This strategy enables sensitivities comparable to those typically obtained in plasmonic sensors while maintaining a large resonance amplitude due to the low-loss dielectric platform. Our results demonstrate that near-field engineering in low-$Q$ dielectric metasurfaces provides an effective pathway to combine high sensitivity with robust and high-contrast spectral responses without relying on ultra-high-$Q$ resonances. In Section 3, the numerical analysis is experimentally verified with ethanol and water acting as analytes, showing the operating principle in analytes with low and moderate losses. 

\section{Near-field engineering and mode hybridization}

\begin{figure}[t]
    \centering
    \includegraphics[width=0.8\linewidth]{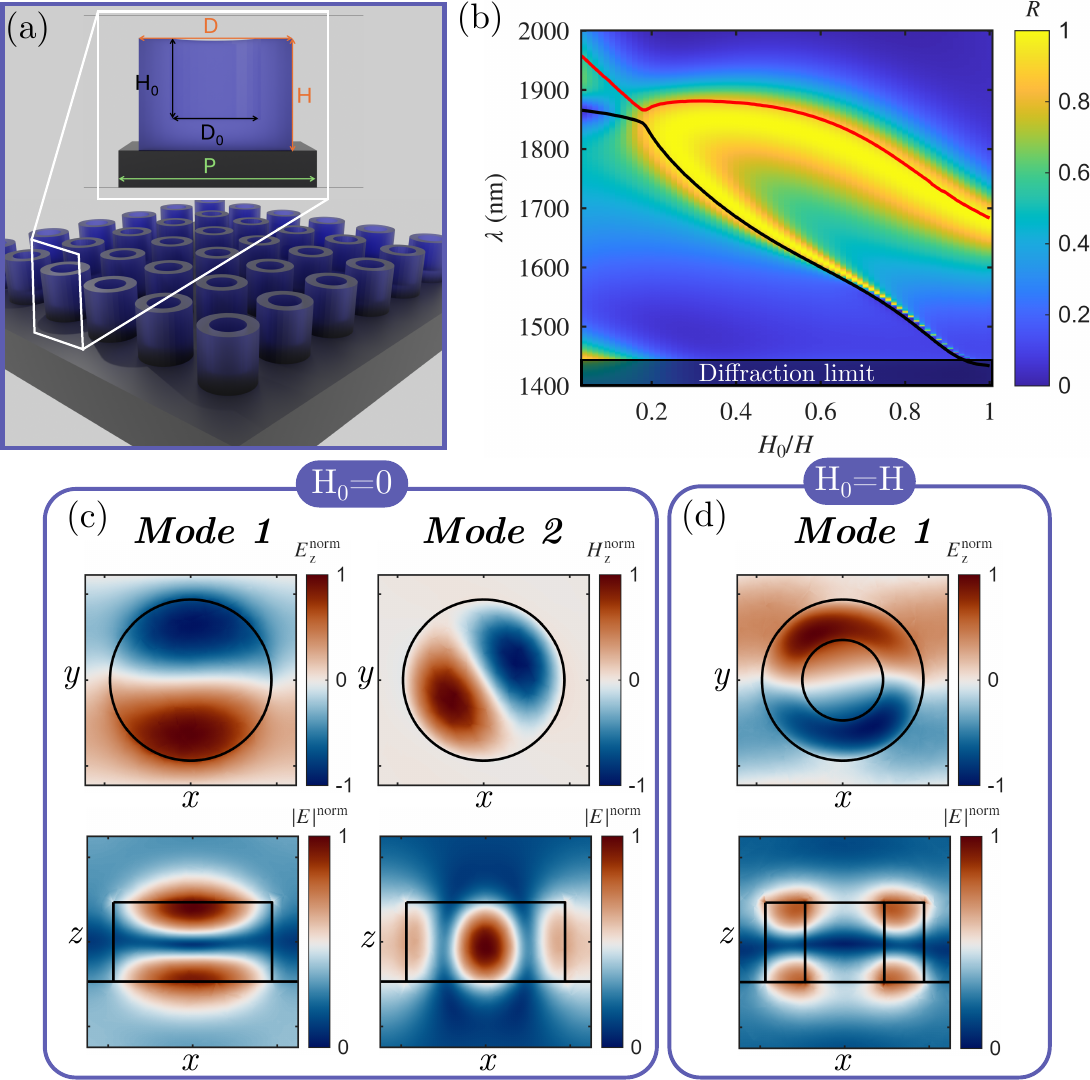}
    \caption{\textbf{Resonant properties of the metasurface.}  (a) Scheme of the metasurface under study and definition of the main geometrical parameters.(b) Reflection spectra of the structure as a function of the $H/H_0$ ratio (the red and black curves show the eigenmode wavelength of mode 1 and mode 2. (c) Electric field distribution of the electric and magnetic dipole modes when $H_0=0$. (d) Electric field distribution of the electric and magnetic dipole modes when $H_0=H$. Geometrical parameters: $H=375$ nm, $D=750$ nm, $P=1000$ nm, and $D_0=D/2$.}
    \label{fig:Panel1}
\end{figure}

In this section, we present a detailed analysis of the field concentration and consequently the density in an all-dielectric structure supporting both electric and magnetic modes, and study the effect of mode hybridization on the sensitivity features. To this end, we use a metasurface consisting of a square array of Si-disks of diameter $D=750 \; \rm nm$, height $H=375 \; \rm nm$, and the distance between disks is $P=1000 \; \rm nm$ on top of a $\rm SiO_2$  substrate.
To study the effect of mode hybridization, we introduce a perturbation in the cylinder geometry. More specifically, a cylindrical hole in the center of diameter $D_0=375 \; \rm nm$ and height $H_0$. Figure \ref{fig:Panel1} (a) shows a schematic view of this structure and its unit cell.

When $H_0 = 0$, the metasurface supports the excitation of both electric and magnetic dipolar Mie-type resonances. Using an eigenmode analysis of the structure, the quality factors are calculated - the electric dipolar resonance reaches a quality factor of approximately $Q \approx 37$, whereas the magnetic dipolar resonance presents a significantly lower value of $Q \approx 7$. The electric field distribution of both electric and magnetic modes is represented in Figure~\ref{fig:Panel1} (c). Due to their different field distributions, the overlap between the electromagnetic field and the surrounding analyte can be quantified through the field–analyte coupling coefficients and the corresponding estimated sensitivity defined in Equation~\ref{eq:1}. A summary of the main parameters of these resonances, including their resonant wavelength, quality factors, sensitivity, and the parameter $f$, is shown in Table~\ref{tab:Table1}.

When we introduce the perturbation in the structure, $H_0 > 0$,  the mirror symmetry of the system is progressively broken. As a result, the originally weakly coupled electric and magnetic dipolar modes begin to interact and hybridize. This interaction gives rise to the phenomenon known as {bianisotropy}, which can be interpreted as an electromagnetic coupling between electric and magnetic responses~\cite{asadchy2018bianisotropic}. Figure~\ref{fig:Panel1}(b) shows the reflectance, $R=|r|^2$, being $r$ the reflection coefficient, spectra of the structure for different values of the perturbation depth $H_0$, where the spectral evolution of the two modes can be observed. As the perturbation increases, the resonances exhibit a clear wavelength dependence associated with the hybridization process between the electric and magnetic dipolar modes. The emergence of this coupling locally modifies the distribution of the electromagnetic near field, thereby altering the overlap between the resonant field and the surrounding analyte region. Consequently, the field-analyte coupling integral is also modified.

In the limiting case where $H_0 = H$, the perturbation fully perforates the resonator, and the bianisotropic response becomes mainly mediated by the presence of the substrate rather than by the internal symmetry breaking within the dielectric resonator itself. Nevertheless, the resulting resonant modes exhibit electric field distributions that are significantly different from those obtained in the unperturbed case ($H_0 = 0$). In particular, the characteristic field patterns associated with purely electric and magnetic dipolar Mie resonances are no longer clearly distinguishable. Instead, the modes present hybridized field distributions that differ from the typical dipolar signatures commonly observed in dielectric nanodisks supporting Mie-type resonances. Although the bianisotropic coupling is weaker in this configuration compared to the intermediate perturbation regimes, the modification of the field distribution remains substantial and directly impacts the field overlap with the surrounding analyte. The corresponding field patterns are shown in Figure~\ref{fig:Panel1}(d), while the main resonant properties and the estimated sensitivity of these modes are summarized in Table~\ref{tab:Table1}. The sensitivities shown in the fourth column of Table~\ref{tab:Table1} have been calculated by eigenmode analysis. The resonance displacement is calculated by comparing the structure surrounded by air ($n=1$) or ethanol ($n=1.346$). Conversely, the fifth column is obtained by calculating Equation~\ref{eq:1} for the structure surrounded by air. The near-field volume in Eq. (1) is defined considering two periods in the vertical direction of the unit cell, centred at the origin and with volume $V_{\rm d}=[P, P, 2P]$, and the analyte volume is defined as $V_{\rm a}=[P, P, 2H]$, directly placed on top of the silica substrate and excluding the resonant silicon structure.

\begin{table}[h]
\centering
\begin{tabular}{c |l c c c c c}
\toprule
& & Mode & $\lambda$ (nm) & Q-factor 
& \makecell{$S$ (nm/RIU) \\ Eigenmode}
& \makecell{$f$ \\ Eq.\,(1)} \\
\midrule

\multirow{6}{*}{\rotatebox{90}{\textbf{D.I}}}
& \multirow{2}{*}{$H_0=0$}
& 1 (ED) & 1977 & 7 & 37 & 0.094\\
& 
& 2 (MD) & 1866 & 37 & 175 & 0.089\\
\cmidrule{2-7}

& \multirow{2}{*}{$H_0=0.2H$}
& 1 (Hybrid-ED) & 1869 & 13 & 158 & 0.092\\
& 
& 2 (Hybrid-MD) & 1830 & 13 & 124 & 0.095\\
\cmidrule{2-7}

& \multirow{2}{*}{$H_0=H$}
& \cellcolor{rowblue}1 
& \cellcolor{rowblue}1683
& \cellcolor{rowblue}13
& \cellcolor{rowblue}282
& \cellcolor{rowblue}0.166\\
& 
& - & - & - & - & - \\
\midrule

\multirow{2}{*}{\rotatebox{90}{\textbf{D.II}}} 
& \multirow{2}{*}{\makecell[l]{$H_0=0$ \\ $D=460\,\mathrm{nm}$}}
& \cellcolor{rowblue}1
& \cellcolor{rowblue}1635
& \cellcolor{rowblue}16
& \cellcolor{rowblue}120
& \cellcolor{rowblue}0.102 \\
& 
& - & - & - & - & -\\
\bottomrule
\end{tabular}
\caption{Sensitivity analysis for different configurations of the structure base on eigenmode analysis. Geometrical parameters: $H=375$ nm,  $P=1000$ nm, and (D.I) $D=750$ nm, $D_0=D/2$, (D.II) $D=460$ nm, $D_0=0$.}
\label{tab:Table1}
\end{table}

\begin{figure}[t]
    \centering
\includegraphics[width=0.8\linewidth]{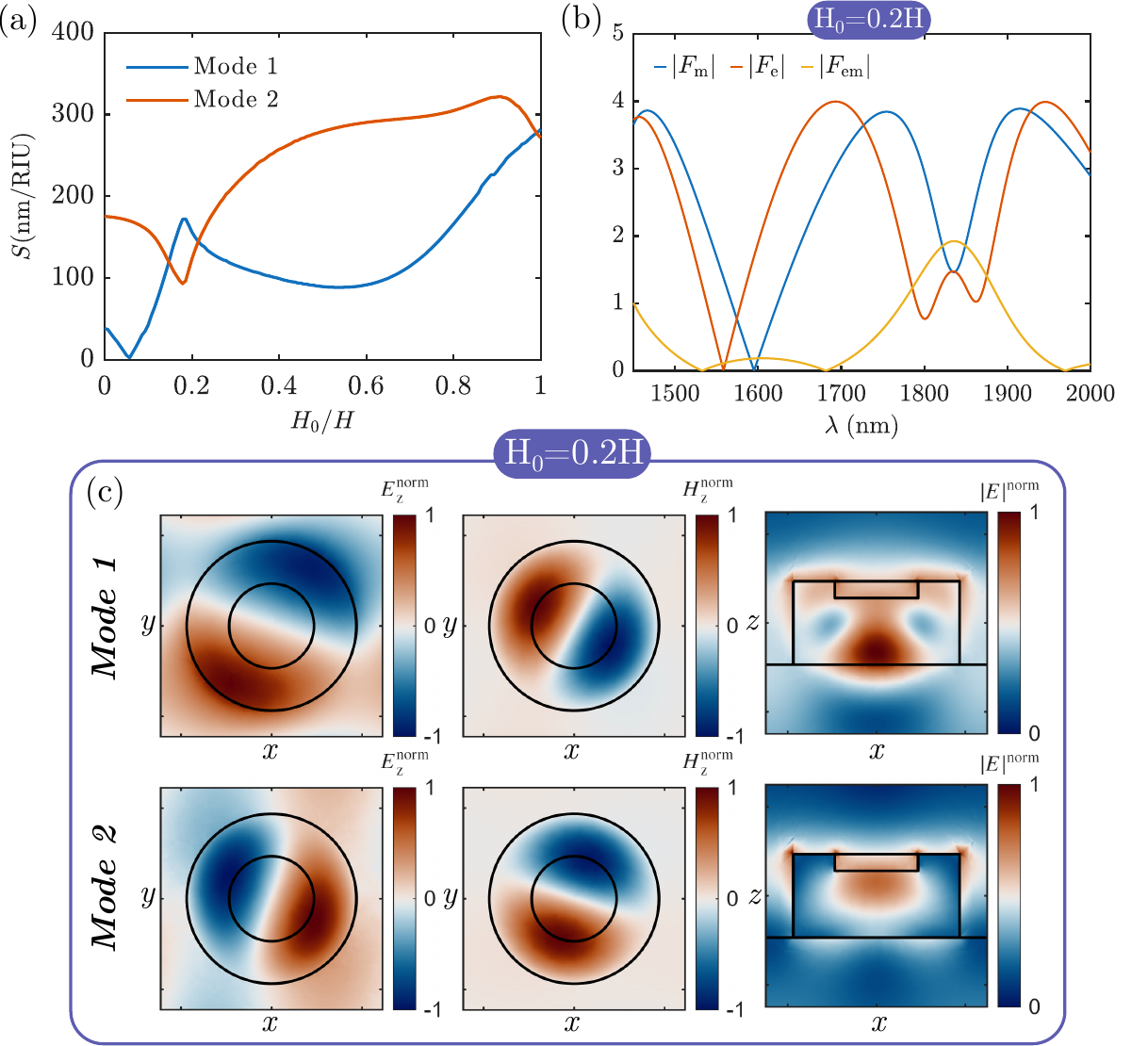}
    \caption{\textbf{Bianisotropy effect in sensitivity.} (a) Analysis of the sensibility for different values of \textbf{$H_0$}  (b) Normalized electric, magnetic, and electromagnetic polarizabilities when $H_0=0.2H$. (c) Normalized electric and magnetic field patterns illustrate the hybrid nature of the resonances and the simultaneous excitation of electric and magnetic responses within the meta-atom. }
    \label{fig: Panel2}
\end{figure}

Interestingly, the evolution of the resonant modes as a function of the perturbation parameter $H_0$, shown in Figure~\ref{fig:Panel1}(b), reveals the emergence of an anticrossing behavior between the two modes. In particular, when the perturbation reaches $H_0 = 0.2H$, a clear mode hybridization occurs, leading to the characteristic avoided crossing in the dispersion of the resonances. This behavior indicates strong electromagnetic coupling between the electric and magnetic modes. At this point, the two hybridized modes present opposite responses in terms of sensitivity: while Mode 1 has a local maximum of the sensitivity, Mode 2 presents a minimum. These behaviors are observed in the analysis shown in Figure~\ref{fig: Panel2}(a) for different values of $H_0/H$. 

For a deeper analysis of this phenomenon, we can study the polarizabilities of the structure. We define the following normalized polarizability terms:
\begin{equation}
    F_{\rm m} =\frac{2\omega}{S_{\rm A} \eta}\alpha^{\rm mm}_{\rm xx}=\frac{2\omega}{S_{\rm A}\eta}\alpha^{\rm mm}_{\rm yy}=-j(r^++r^--2t+2) 
\end{equation}
\begin{equation}
    F_{\rm e}=\frac{2\omega \eta}{S_{\rm A}}\alpha^{\rm ee}_{\rm xx}=\frac{2\omega \eta}{S_{\rm A}}\alpha^{\rm ee}_{\rm yy}=j(r^++r^-+2t-2)
\end{equation}
\begin{equation}
    F_{\rm em}=\frac{2\omega }{S_{\rm A}}\alpha^{\rm em}_{\rm xy}=-\frac{2\omega }{S_{\rm A}}\alpha^{\rm em}_{\rm yx}=-j(r^+-r^-)
\end{equation}
where $\alpha^{\rm ee}_{ij}$, $\alpha^{\rm mm}_{ij}$, and $\alpha^{\rm em}_{ij}$ are the terms of the electric, magnetic and magneto-electric polarizability matrices, $S_{\rm A}$ is the area of the unit cell, $\eta$ is the impedance of the medium, $\omega$ is the angular frequency, $t$ is the transmission coefficient, $r^+$ is the reflection coefficient when illuminating the metasurface from the top, and $r^-$ is the reflection coefficient when illuminating the metasurface from the bottom. 

Figure~\ref{fig: Panel2}(b) shows the polarizability study when $H_0=0.2H$. Remarkably, this specific value of $H_0$ corresponds to the condition of a balanced omega particle, where the electric and magnetic responses of the meta-atom become comparable, resulting in a symmetric modal interaction and the observed anticrossing in the spectrum. Figure~2(c) shows cross-sectional maps of the electric field magnitude for the structure at the configuration $H_0 = 0.2H$. For this case, the field–analyte overlap integral defined in Equation~\ref{eq:1} was computed, and the corresponding values are reported in Table~\ref{tab:Table1}. 

However, despite the presence of a local maximum in sensitivity around $H_0 = 0.2H$, the structure exhibits an even higher sensitivity when $H_0 = H$, i.e., when the central hole extends through the entire height of the cylinder. In this configuration, although the electromagnetic coupling coefficient decreases, the field–analyte increases, as observed in both Figure~2(a) and Table~1. As a result, the overall sensitivity of the resonance becomes larger. Furthermore, this geometry presents a simpler fabrication process, since it can be realized in a single lithography step. For these reasons, in the following,
we focus our analysis on this configuration and compare its performance with that of a conventional nanodisk without a central hole.

\begin{figure}[t]
    \centering
\includegraphics[width=1\linewidth]{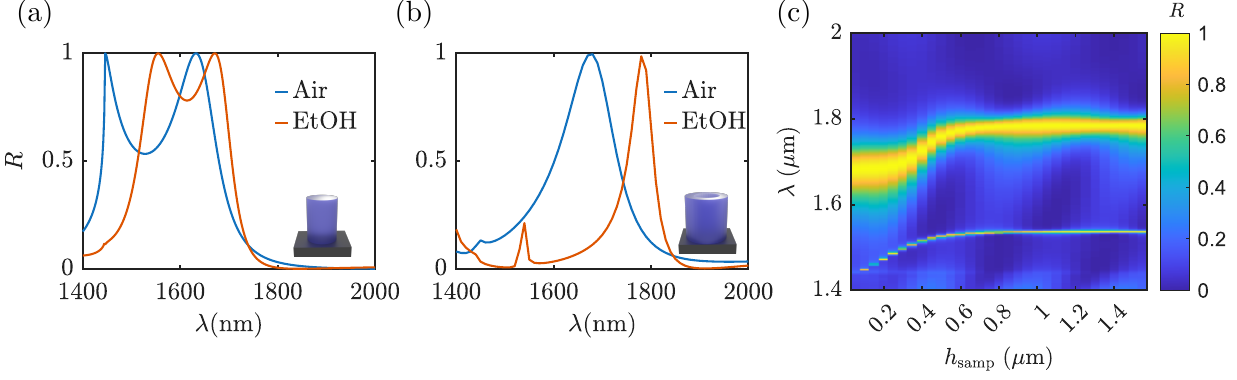}
    \caption{\textbf{Spectral response of the metasurface for different surrounding media.} 
(a) Reflection spectra of a reference metasurface composed of Si nanodisks with diameter $D= 460$ nm, height $H=375$ nm, and period $P= 1000$ nm placed on a $\mathrm{SiO_2}$ substrate, when the surrounding medium changes from air to ethanol (EtOH).  (b) Reflection spectra of the metasurface structure with $D= 750$ nm, $H=375$ nm, and $P= 1000$ nm, including a single central perforation of diameter $D_0=375$ nm and depth $H_0=375$ nm, showing the spectral response for air and ethanol as surrounding media. (c) Reflection spectra for a parametric sweep of the analyte layer thickness deposited on top of the perforated resonators metasurface. }
    \label{fig: Panel3}
\end{figure}

For a better comparison of the advantages introduced by the near-field engineering, we also compare the engineer mode with $H_0=H$ with a nanodisk without holes with a similar Q factor. Figure~\ref{fig: Panel3}(a) shows the reflection spectra and field profile for the non-perforated structure, with $H=375$ nm, $H_0=0$ nm, $D=460$ nm, and $P=1000$ nm, while Figure~\ref{fig: Panel3} (b) shows the same for a fully perforated disks structure with $H=H_0= 375$ nm, $D= 750$ nm, $D_0 = 375$ nm, and $P=1000$ nm, for two different refractive indices. The diameter of the disks was adjusted relative to that of the resonators with holes to obtain modes at similar wavelengths and with comparable Q values. Both structures are compared in Table 1, where we can see that both modes have a higher sensitivity for the structure composed of resonators with holes, with mode 2 showing more than double the sensitivity.

Figure~\ref{fig: Panel3}(c) shows a parametric sweep of the thickness of the analyte layer placed on top of the metasurface for the fully perforated structure. The analyte thickness is varied from 0 nm up to 1500 nm in order to evaluate its influence on the resonant response of the structure. It can be observed that for thickness values below approximately 200 nm, the resonance wavelengths remain almost unchanged, i.e.,  the sample is not detectable, and the resonance is centered around the same wavelength as in the case of no sample. 
A noticeable shift appears for intermediate thicknesses between roughly 250 nm and 600 nm, where the interaction between the analyte and the evanescent field of the metasurface becomes stronger. For larger thicknesses, the resonance position stabilizes, and no further significant changes are observed. This behavior indicates that only the portion of analyte located within the near-field region contributes effectively to the sensing response, while additional analyte beyond this interaction volume does not further modify the resonance.

\section{Metasurface fabrication and characterization}

\begin{figure}[t]
    \centering
    \includegraphics[width=0.8\linewidth]{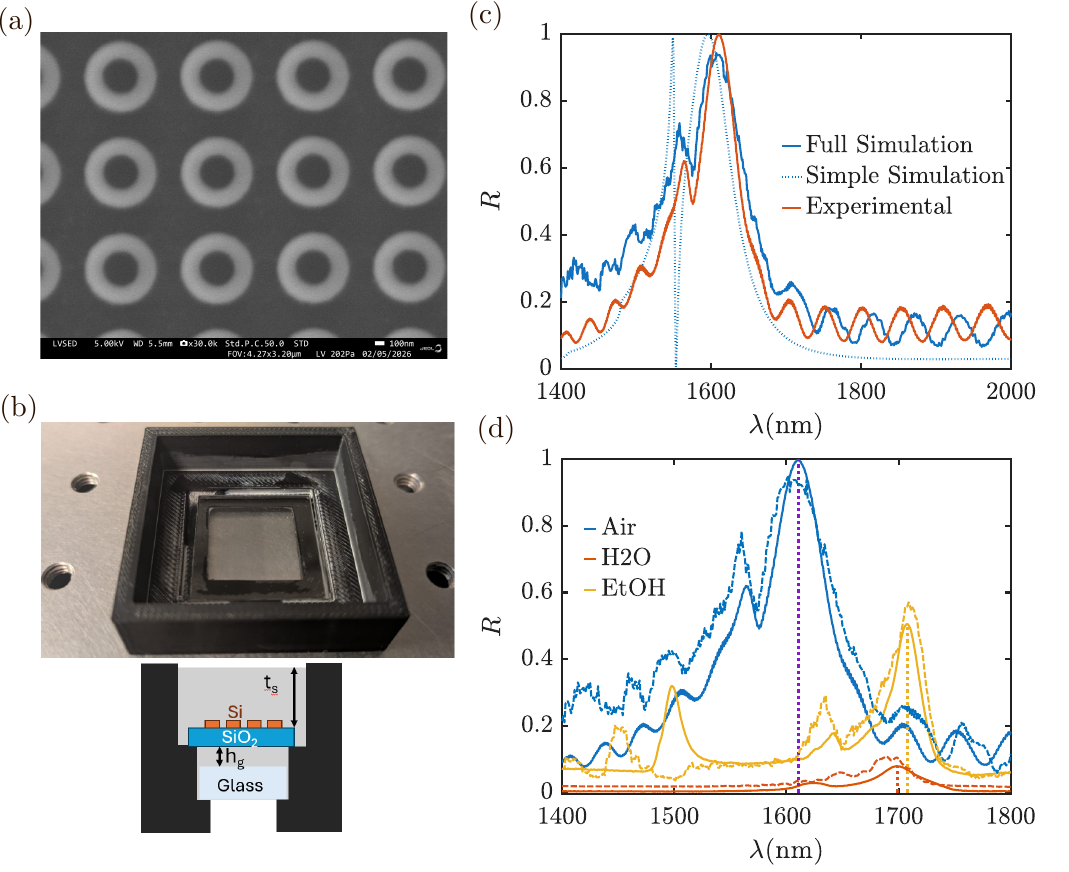}
   \caption{\textbf{Experimental demonstration of the sensing concept.} 
(a)  A scanning electron microscope (SEM) image of the fabricated metasurface is also shown. (b) Schematic of the measurement setup based on FTIR reflection spectroscopy. The inset shows a cross-sectional diagram of the sample placed inside the custom cuvette, including the analyte layer, the Si metasurface on a SiO$_2$ substrate, and the glass window. (c) Comparison between the experimentally measured reflection spectrum of the metasurface without analyte and two numerical simulations considering an infinite substrate and including the finite thickness of the substrate, showing improved agreement with the experiment. (d) Reflection spectra for different surrounding analytes (air, ethanol, and water). The full simulation in (c) and the simulations in (d) were post-processed using a smoothing filter.}
\label{fig: Panel4}
\end{figure}

The resonant structures were fabricated on a 365 nm thick layer of amorphous silicon deposited on a glass substrate using plasma-enhanced chemical vapor deposition (PECVD). The PECVD deposition of the amorphous silicon was carried out using a CENTURA 5200 system (Applied Materials), with silane (SiH$_4$) and hydrogen (H$_2$) as precursor gases. The fabrication process was based on electron-beam direct writing performed on a 100 nm thick hydrogen silsesquioxane (HSQ) resist layer. The electron-beam exposure was carried out using a Raith150 system and optimized to achieve the desired feature dimensions. The exposure parameters included an acceleration voltage of 30 keV and an aperture size of 30 µm. After exposure, the HSQ resist was developed using tetramethylammonium hydroxide (TMAH). The resulting resist patterns were then transferred into the samples through an optimized inductively coupled plasma reactive ion etching (ICP-RIE) process using fluorine-based gases (CF4 and C4F8). Figure~\ref{fig: Panel4} (a) shows a scanning electron microscope (SEM) photograph of the metasurface.  The nominal fabrication parameters were $D= 750$ nm, $D_0 = 375$ nm, and $P=1000$ nm, although due to manufacturing tolerances, the fabricated parameters were deviate from those. The effective parameters were obtained from fitting the experimental data to the simulated spectra for different parameter sets. The best fitting results were obtained for $D= 700$ nm, $D_0 = 400$ nm, and $P=1000$ nm.

To measure the sensing performance of the metasurface, it was placed inside a cuvette made for measuring liquid samples. The cuvette is made of PLA, with a glass window at the bottom. The cuvette has 3 mm thick walls, a total height of 13 mm (with an internal depth of 10 mm), and a side length of 46 mm (corresponding to an internal side length of 40 mm). The inset in Figure~\ref{fig: Panel4} (b) shows the placement of the glass window and the metasurface in the cuvette.
The reflection spectra of the metasurface were taken using a Vertex 80 FTIR system with a Hyperion microscope, and using a custom objective that allows near-normal incidence illumination. Figure~\ref{fig: Panel4} (c) shows the spectrum of the metasurface with the empty cuvette. The resonance shows two peaks, one of them appearing due to non-normal incidence, and out of the resonance, the spectrum presents Fabry-Perot oscillations created by the small air gap $h_{\rm g}$ between the substrate of the metasurface and the glass window. From this effect, we estimate the gap height to be approximately $30 \mu \rm m$.
Figure~\ref{fig: Panel4} (c) compares the obtained spectrum measuring with an empty cuvette to two simulations: one being a simple simulation with just the metasurface with an infinite substrate, and the other a full that takes into account the finite size of both the substrate and the glass window of the cuvette, as well as the small gap between them $h_{\rm g}$. The large thickness of these elements leads to a very thin Fabry-Perot effect that cannot be experimentally measured, but appears in the simulations. We have smoothed the data from the full simulation to better emulate the spectra obtained from measurements. For both simulations, we used an incidence angle of $2^{\rm o}$ to model the small aperture angle of the custom-made objective, making the second peak appear at a similar wavelength to the one seen in the experiment.

To test the sensing performance of this metasurface, we performed an experiment that consisted of partially filling the cuvette with the sample liquid using a pipette until the whole metasurface was covered and measuring the reflection spectrum. We used two different liquids for this: water, with $n=1.307$ and $k=8\cdot 10^{-5}$ at 1700nm~\cite{segelstein1981complex}, and ethanol, with $n=1.346$ and $k=1.13\cdot 10^{-4}$ at 1700nm~\cite{myers2018accurate} ($n$ and $k$ at the whole wavelength range can be found in the Supplementary). The depth of liquid on top of the metasurface $h_{\rm samp}$ is estimated by fitting the amplitude of the resonance peak to simulations for different $h_{\rm samp}$ values. Although a larger value of sample depth leads to a lower reflection peak due to higher overall absorption, the absorption due to interaction of the localized field near the metasurface with the sample will be the same, meaning there will not be important changes to its Q or sensitivity.
Figure~\ref{fig: Panel4} (d) shows the results of the measurements compared to simulations. The $h_{\rm samp}$ values in the simulations (0.3 mm for EtOH and 2 mm for $\rm H_2O$) were chosen to fit the amplitude of the resonances. We can see a shift of 9 nm in the reflection maximum between water and ethanol, which means a sensitivity of $S=281 \rm \; nm/RIU$, similar to the one obtained from the simulations in Table 1.

\section{Conclusions}

In this work, we have demonstrated that the sensing performance of dielectric metasurfaces can be significantly enhanced through near-field engineering of low-$Q$ resonant modes. By introducing a controlled geometric perturbation in silicon nanodisk resonators, we enabled the hybridization of electric and magnetic dipolar Mie resonances, which modifies the spatial distribution of the electromagnetic field and increases the overlap between the resonant mode and the surrounding analyte. As predicted by the theoretical sensitivity expression, this redistribution of the near field leads to an increase in the fraction of electromagnetic energy located within the sensing region and therefore enhances the refractive-index sensitivity.

A parametric analysis of the perturbation depth revealed the presence of a local sensitivity maximum associated with strong electric–magnetic mode hybridization around $H_0=0.2H$. However, the highest overall sensitivity was obtained when the resonator is fully perforated ($H_0=H$), where the modified field distribution increases the field–analyte overlap despite a reduction in the electromagnetic coupling coefficient. In this configuration, the metasurface exhibits sensitivities exceeding those of comparable nanodisk resonators with similar quality factors, highlighting the effectiveness of the proposed near-field engineering strategy.

The sensing mechanism was further validated experimentally using FTIR reflection spectroscopy. The measured spectra show good agreement with numerical simulations and demonstrate a clear resonance shift when the surrounding medium changes from water to ethanol, corresponding to an experimental sensitivity of approximately $281\,\mathrm{nm/RIU}$. These results confirm that low-loss dielectric metasurfaces can achieve sensing performances comparable to those typically associated with plasmonic sensors while maintaining high resonance amplitudes due to the absence of Ohmic losses.

Overall, our results show that optimizing the spatial distribution of the electromagnetic near field provides an effective pathway to improve the sensing capabilities of metasurface platforms without relying on ultra-high-$Q$ resonances. This approach offers a robust and experimentally accessible strategy for the design of high-performance optical sensors and may be extended to other metasurface geometries and material platforms for chemical and biological detection.

\section{Supplementary}

Amorphous silicon was optically characterized by ellipsometry. The values for its refractive index were used for simulations and are plotted in Figure S1.

\begin{figure}[H]
    \centering
    \includegraphics[width=0.5\linewidth]{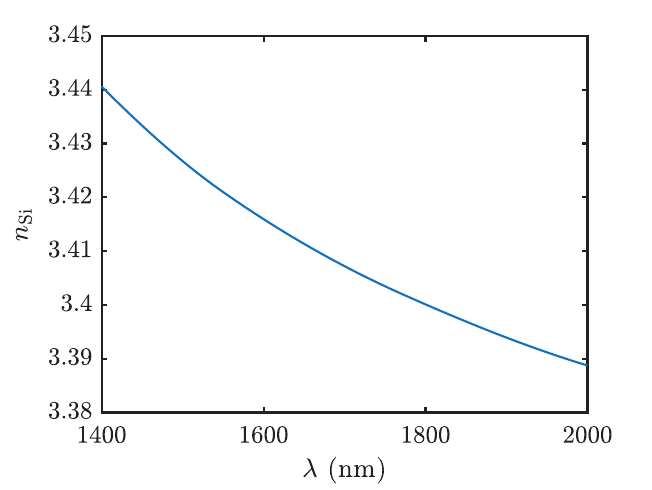}
    \caption*{Figure S1: Refractive index of Silicon as a function of the wavelength.}
    \label{fig:placeholder}
\end{figure}

Water and ethanol values of refractive index and extinction coefficient (from~\cite{segelstein1981complex} and~\cite{myers2018accurate} respectively) are plotted in Figure S2.

\begin{figure}[H]
    \centering
    \includegraphics[width=1\linewidth]{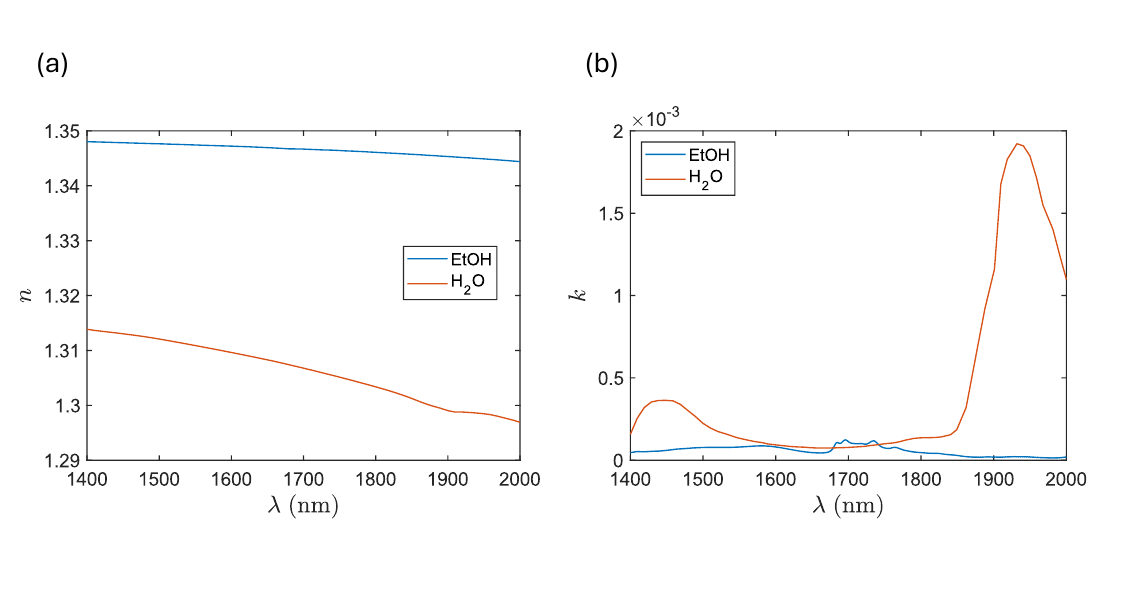}
    \caption*{Figure S2: (a) Refractive index and (b) extinction coefficient of $\rm H_2O$ and $\rm EtOH$ as a function of the wavelength.}
    \label{fig:placeholder2}
\end{figure}

In Figure S3, we show the values of $f$ for both modes as a function of the hole height $H_0$. The fixed volumes used are defined as $V_a=[P,P,H]$ and $V_d=[P,P,2H]$
\begin{figure}[H]
    \centering
    \includegraphics[width=0.5\linewidth]{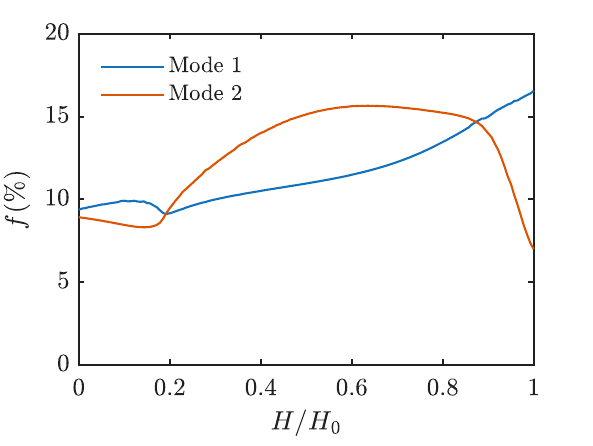}
    \caption*{Figure S3: $f$ as a function of the hole height for the two studied modes. Mode 1 shows a similar trend to that of its sensitivity shown in Figure 2(a) of the main text.}
    \label{fig:placeholder}
\end{figure}

\newpage

\section*{Funding}
Generalitat Valenciana PROMETEO Program (CIPROM/2022/14), Spanish National Research Council (grant no. PID2024-162261NB-I00, CNS2024-154715), and the Universitat Polit\`ecnica de Val\`encia (grant no. PAID-01-23 and POLISABIO2024-P18).

\section*{Disclosures.} The authors declare no conflicts of interest. 

\section*{Data availability.} Data underlying the results presented in this paper are not publicly available at this time but may be obtained from the authors upon reasonable request.

\printbibliography

\end{document}